\documentclass[twocolumn,floats,,amsmath,amssymb]{revtex4}
\usepackage{graphicx}
\usepackage{dcolumn}
\usepackage{bm}

\begin{document}
\draft

\title{Theory of adiabatic Hexaamminecobalt-Self-Exchange}

\author{R. G. Endres, M. X. LaBute, D. L. Cox}
\address{Department of Physics, University of California, Davis, CA 95616}

\date{\today}

\begin{abstract}
We have reexamined the thermally induced $Co(NH_3)_6^{2+/3+}$ [Co(II/III)] 
redox reaction using the first principles density-functional-theory method, 
semiclassical Marcus theory, and 
known charge transfer parameters. We confirm a previously suggested mechanism 
involving excited state (${}^2E_g$) of Co(II) which
becomes lower than the ground state (${}^4T_{1g}$) in the transition state region. This lowers the transition state barrier 
considerably by $\sim 6.9 kcal/mol$ and leads to a spin-allowed and 
adiabatic electron exchange process. Our calculations are consistent 
with previous experimental results regarding the spin-excitation energy (${}^3T_{1g}$) of 
Co(III), and the fact that an optical absorption peak (${}^2E_g$) of the Co(II) species 
could not be found experimentally. Our rate is of order $6\cdot 10^{-3}(Ms)^{-1}$ and hence 
2 orders of magnitude faster than determined previously by experiments. 

\end{abstract}

\maketitle

\section{Introduction}

The experimental rate determination and theoretical 
understanding of the $Co(NH3)6^{2+/3+}$ redox reaction in aqueous solution 
\begin{eqnarray}
  &&\!\!\!\!\!\!\!\!Co(NH_3)_6^{3+}+Co(NH_3)_6^{2+}\stackrel{k}{\longrightarrow}\nonumber \\
  &&\qquad Co(NH_3)_6^{2+}+ Co(NH_3)_6^{3+}\label{redox}
\end{eqnarray}
has been a great intellectual challenge for several decades. Despite substantial
effort, the mechanism of the rate is still an unsolved problem. Is the reaction 
spin-forbidden and diabatic or are spin-excited states thermally accessible which
would possibly lead to a spin-allowed, adiabatic reaction?
 
Experimental studies in the early 1960s suggested 
an extremely slow rate of $k=1.6 \cdot 10^{-10}(Ms)^{-1}$ \cite{stranks} 
at 64.5 C and 1 M ionic strength using radiocobalt ${}^{60}$C as a tracer. 
Unfortunately, side reactions involving hydrolysis of the complexes, which contribute 
to the rate, have not correctly been taken into account. 
In the 1980s a detailed analysis of previous data using the 
Marcus correlation\cite{marcus63} led to a much higher estimate of
$1\cdot 10^{-5}(Ms)^{-1}$ \cite{geselowitz1}, and subsequent experiments 
labeling the ammonia ligands with ${}^{15}$N 
(which can be assayed by NMR of the ammine protons and make it possible 
to trace side reactions)  gave about $6 \cdot 10^{-6}(Ms)^{-1}$ at 40 C 
and 2.5 M ionic strength \cite{geselowitz2}. An estimate of the rate at 
normal conditions (25 C and 1M ionic strength) gives 
$\lesssim 5\cdot 10^{-7}(Ms)^{-1}$ \cite{brunschwig,sutin1,sutin2,newton91}.
These results can be expected to be rather trustworthy, because 
they are, as expected, similar to the rate of a 
closely related and well understood system, $Coen_3^{3+/2+}$, which
has a rate of $5.2 - 7.7\cdot 10^{-5}(Ms)^{-1}$\cite{Coen}.
Furthermore, an early but less influential measurement of 
$Co(NH_3)_6^{2+/3+}$ in liquid $NH_3$ 
(and hence not suffering from hydrolysic side reactions) 
indicated a similar rate of order $10^{-5} (Ms)^{-1}$\cite{liquidNH3}.

Briefly, previous results of theory mainly by Buhks {\it et al} (1978) \cite{buhks} and 
Newton (1986,1991) \cite{newton86,newton91} indicate that the reaction involves the ground state
species low-spin $Co(NH_3)_6^{3+}$, ${}^1A_{1g}$ [Co(III,S=0)], 
and high-spin $Co(NH_3)_6^{2+}$, ${}^4T_{1g}$ [Co(II,S=3/2)]. 
The spin-excited states, ${}^3T_{1g}$ [Co(III,S=1] and ${}^2E_g$ [Co(II,S=1/2], 
were to high in energy to be thermally populated. 
This led to a spin-forbidden (only possible by weak spin-orbit coupling), 
diabatic reaction with a rate constant of about 4 orders of magnitude 
too small compared to experiment. However, work by Larsson {\it et al} (1985) 
stressed that the ground state-excited state energy separations
at the transition state have to be considered, not at the equilibrium geometries. Including Jahn-Teller (JT)
stabilization energies for the excited states, this led in particular to a substantial energetic 
lowering of Co(II,S=1/2) relative to Co(II,S=3/2). Although this made the reaction spin-allowed, their work
suffered from inconsistent data from different electronic structure codes. Furthermore, the rate constant 
was not determined.

In this paper we show that the thermal activation barrier for the
spin-allowed reaction between Co(III,S=0) and Co(II,S=1/2) [total spin S in units of $\hbar$]
is drastically lowered ($\sim 6.9 kcal/mol$)
compared to the spin-forbidden reaction between Co(III,S=0) and Co(II,S=3/2).
Unlike previously believed, the JT-distorted spin-excited Co(II,S=1/2) and the ground state Co(II,S=3/2) are 
near degenerate. Hence the reaction is spin-allowed and adiabatic. 
Our results are based on the density functional theory (DFT)\cite{kohn} 
code SIESTA\cite{siesta} which we use to 
calculate spin-excited states of isocharge molecular species and potential energy surfaces (PESs).
Since DFT is a ground state theory, excitation energies obtained 
from ground state energy differences of species with different total spin 
are expected to be rather good. The DFT method has the advantage over Hartree-Fock methods that
certain correlations are already build in.  
From the PESs we determine the activation barrier and
provide a new estimate of the rate constant utilizing previously estimated quantities 
like the $e_g$ - electronic coupling $H_{DA}={}_D\!\!<\!\!e_g|H|e_g\!\!>_A$ between donor D and acceptor A
molecular species, the outer-sphere contribution to the reaction barrier 
$E^\dagger_{out}$, and the preequilibrium constant $K_{equ}$. 
Although our excitation energy of Co(III), $\Delta E^{3+}|_{equ}=13,120 \text{cm}^{-1}$, 
is in rather good agreement with the experimental value, we deemphasize the absolute numerical values of
excitation energies. Instead we emphasize trends established from DFT calculations exploiting
that DFT usually stabilizes high-spin over low-spin states.  
If Co(II) has a low-spin ground state, this would also explain why the optical absorption 
${}^2E_g\leftarrow{}^4T_{1g}$ has not been observed. 

The paper is organized as follows. In section \ref{section2} we review further experimental
and previous theoretical efforts in more detail. In section \ref{section3} we introduce the DFT code SIESTA 
and explain our computational methods (section \ref{subsection3.1}), as well as present the 
calculation of the PESs (section \ref{subsection3.2}). In section \ref{section4} we provide a 
new estimate of the spin-allowed hexaammine self-exchange reaction based on our insights from 
the PESs and discuss our results. Finally, we summarize in section \ref{section5}.

\section{\label{section2}Review of previous efforts}

In this section further experimental facts about the single complexes and a more detailed 
review of past theoretical efforts are presented.

It is well accepted that $Co(NH_3)_6^{3+}$ [Co(III)] is a 
stable low-spin (S=0) compound, while it is {\it assumed} that  
$Co(NH_3)_6^{2+}$ [Co(II)] is high-spin (S=3/2).
There are several reasons for this. First it is known from 
${}^{59}$ Co NMR studies that the related system, $Co(H_2O)_6^{3+}$, 
shows no sign of exchange with paramagnetic species\cite{navon} and hence is low-spin. 
This is also expected to be true for Co(III), since it has an even
larger ligand-field favoring low-spin.
X-ray diffraction data of related crystals shows further a drastic difference of the 
Co-N bond distances ($\approx 0.22 \AA$ \cite{kime,brunschwig}) 
between Co(II) and Co(III). 
From this it was concluded that the ligand-field of Co(II) must be much smaller
resulting in high-spin.
Additionally, the optical excitation spectrum of Co(III) could be 
fully characterized including $d-d$ spin-forbidden transitions. This is different than Co(II), 
where it could not be measured successfully. 
It is further experimentally supported that the $Co(NH_3)_6^{2+/3+}$ self-exchange
reaction occurs as outer-sphere. This is due to a rather slow ligand exchange rate
of Co(III). There is considerable thermodynamic and kinetic stability, which
arises from effective $\sigma$-donation into empty $e_g$-shells\cite{richens}.
 
\begin{figure}
\includegraphics[height=8.0cm,angle=0]{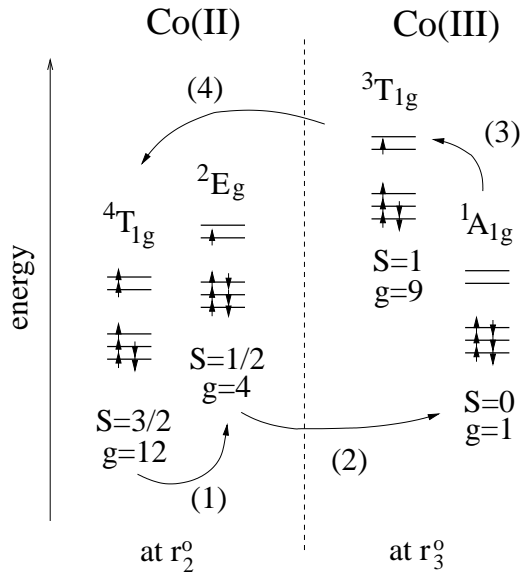}\\
\caption{\label{cartoon}Introduction of the ground state ${}^4T_{1g}$ [${}^1A_{1g}$] 
and first excited state ${}^2E_g$ [${}^3T_{1g}$] of Co(II) [Co(III)], 
their irreducable representations, total spin S [$\hbar$], degeneracy g, and possible 
couplings (1) to (4). The energy ordering corresponds to the 
equilibrium Co-N bond lengths $r_2^o$ and $r_3^o$ of Co(II,S=3/2) and Co(III,S=0), respectively.}
\end{figure}

The theoretical effort is conveniently discussed in the context of the 
separable semi-classical transition-state model\cite{marcus56}
\begin{equation}
  k_{et}=K_{equ}\nu_{eff}\kappa_{el}\Gamma_n\text{exp}(-\beta G^\dagger)\label{k1}.
\end{equation}
In this equation, $K_{equ}$ is the preequilibrium factor 
describing the probability to form a precursor compound, 
$\nu_{eff}$ is an effective nuclear attempt frequency to reach
the transition state (TS), 
$\kappa_{eq}\leq 1$ is the electronic transmission coefficient 
evaluated at the TS and averaged over all possible precursor compounds, 
$\Gamma_n\geq 1$ is the nuclear tunneling factor, and the Boltzmann factor (classical 
Franck-Condon factor) gives the probability to reach the TS with activation free-energy $G^\dagger$ 
($\beta=1/(k_BT)$). For the further discussion we review the 
nomenclature of the important states in figure \ref{cartoon}, i.e. 
their group theoretical representations, total spin S, degeneracy g, and possible 
couplings (1) to (4). The couplings (1) and (3) are due to
spin-orbit coupling or thermal population at finite temperature T, (2) and (4)
are mainly due to mixing of the electronic $e_g$-orbitals of Co(II) and Co(III) 
mediated by their ligands when a precursor state is formed.

It has been known for about 50 years that extraordinarily slow rates can often be
attributed to small Franck-Condon nuclear overlaps. These originate from large changes 
in the metal-ligand bond length in the first coordination sphere\cite{libby}. 
This seemed to apply also to the case 
at hand, if one assumes an adiabatic reaction ($\kappa_{el}\approx 1$)
"neglecting" the spin-forbiddeness. However, a first detailed theoretical study 
by E. Buhks and co-workers using perturbation theory in the weak spin-orbit coupling 
to admix spin-excited states shows that the electronic factor is
very diabatic, $\kappa_{el}\approx 10^{-4}$ \cite{buhks}. 
This makes the rate constant too small. 

Orgel\cite{Orgel} and by Stynes and Ibers\cite{StynesIbers} put out an idea
that the reaction could involve thermally excited Co(II,S=1/2) making it spin-allowed. 
Larsson {\it et al} \cite{larsson} went further and argued that the excited states can become 
much lower in energy near TS. The important 
quantities are the energy difference between first excited state and ground state of 
Co(II) and Co(III) as a function of the Co-N bond length r
\begin{eqnarray}
   \Delta E^{3+}(r)&=&E({}^3T_{1g})-E({}^1A_{1g})\label{dE3}\\
   \Delta E^{2+}(r)&=&E({}^2E_g)-E({}^4T_{1g})\label{dE2}.
\end{eqnarray}
In the case of Co(III) the lowering enhances somewhat the admixture to 
the ground state, while in case of  Co(II), ${}^2E_g$ becomes even lower 
than ${}^1A_{1g}$ circumventing the spin-barrier. Their argument 
is based on Born-Oppenheimer potential energy surfaces (PES) calculated 
both with {\it ab initio} Hartree-Fock (HF) and semi-empirical INDO-CI 
methods including e${}_{g}$ JT effects for both excited states. 
Unfortunately their result was not convincing, since the HF 
calculation predicts the wrong ground state for Co(III), while the 
INDO-CI excitation energy $\Delta E^{2+}|_{equ}=2,000 - 3,000$ cm${}^{-1}$ 
was much lower than the generally believed  
$9,000$ cm${}^{-1}$\cite{buhks}, although there is no experimental evidence 
for this value. There has been no absorption found 
in this region. On the other hand, the excitation energy 
of Co(III) is known from experiment to be 13,700 cm${}^{-1}$\cite{wilson} 
at the Co(III) equilibrium configuration. Since their calculated value 
$\Delta E^{3+}|_{equ}=14,800$ cm${}^{-1}$ is too large by an amount 
$c\approx 1,100$ cm${}^{-1}$, the PES from INDO-CI were corrected according to
\begin{eqnarray}
  \Delta E^{3+}&\rightarrow &\Delta E^{3+}-c\label{c1}\\
  \Delta E^{2+}&\rightarrow &\Delta E^{2+}+c\label{c2}.
\end{eqnarray}
It was noted that the semi-empirical INDO-CI method is generally very 
successful in calculating spectra of transition-metal complexes at fixed 
geometries, but not so in predicting molecular geometries\cite{INDO}. 

In 1991, Newton carried out new {\it ab initio} calculations at both the SCF(UHF) and correlated 
(UMP2) level. Using an empirical correction factor as large as $c=-6,000$ cm${}^{-1}$ 
(HF generally favors high-spin), he estimated the excitation energies at the transition
state to be $\Delta E^{3+}=8,800$ cm${}^{-1}$ and $\Delta E^{2+}=5,300$ cm${}^{-1}$, 
and concluded that thermally excited pathways are not competitive to the 
spin-forbidden ground state pathway at room temperature (RT). 
This left the problem unsolved.

\section{\label{section3}{\it Ab inito} calculations}

\subsection{\label{subsection3.1}Method}

Our results of the inner-sphere activation barrier 
are based on PESs obtained from the fully {\it ab initio} code 
SIESTA\cite{siesta} based on density functional theory (DFT). SIESTA uses 
Troullier-Martins norm-conserving  pseudo potentials\cite{TM} in the 
Kleinman-Bylander form\cite{KB}. For cobalt, we included spin-polarization 
and non-linear core corrections\cite{nlc} to account for a spin-dependent 
exchange splitting and correlation effects between core and valence electrons, 
respectively. Relativistic effects are included for the core electrons in the 
usual scalar-relativistic approximation (mass-velocity and Darwin terms)
and by averaging over spin-orbit coupling terms, while no spin-orbit coupling is 
included for the 4s and 3d valence electrons. This has the computational advantage that
spin remains a good quantum number, while resulting errors of the total energy 
are less crucial when one is interested in total energy differences. 
SIESTA uses a local basis set of pseudo atomic orbitals (PAO) of 
multiple $\zeta$-type. The first-$\zeta$ orbitals 
are produced by the method by Sankey and Niklewski\cite{sankey}, while the 
higher-$\zeta$ orbitals are obtained from the split valence method well known from quantum
chemistry. Polarization orbitals can also be included. 
We used a double-$\zeta$ basis set with polarization orbitals (DZP),
as well as the generalized gradient approximation (GGA) in the version by 
Perdew, Burke and Ernzerhof\cite{PBE} for the exchange-correlation energy functional. 
The charge densities are calculated on a real space grid, where the fineness of the 
grid corresponds to an energy cut-off 80 Ry. 

Before going into the details of how we obtained the PESs plots, we outline 
the main ingredients of our calculations.

\begin{itemize}
\item Our calculations rest on the fixed-spin method within the spin-polarized 
DFT frame work. Two different 
Fermi energies, one for spin-up and one for spin-down, are adjusted in a 
self-consistent way in order to obtain the ground state of a desired total spin.
If a ground state of a certain spin is higher than the ground state
of a different spin, then we know the former as the excited state and the latter one
as the true ground state. Nevertheless, there might be several
states within a spin-manifold which differ in orbital symmetry. 
Since the DFT method is based on the variational principle,
we obtain the state of a system with a certain total spin 
which has the lowest energy\cite{gunnarsson76}. Our
excitation energies are extracted from total energy differences using the well-established 
self-consistent-field method ($\Delta$SCF)\cite{jones}. This gives generally
reliable results for molecules since final state effects are included.

\item The ammines are treated as rigid bodies, i.e. the N-H bond lengths and the 
H-N-H bond angles stay the same throughout all our calculations. This is justified, 
since the vibrational frequencies of covalent bonds (N-H) are about one order of magnitude
higher than the metal-ligand stretching frequencies\cite{sutin1} and 
are considered average values. 

\item We perform geometrical conjugate gradient optimizations 
(CGOs) of both Co(II) and Co(III) with the spin being fixed to a desired value. 
This is used to obtain the
equilibrium geometries (EQGEOs) and the PES. The EQGEOs  
are optimized or relaxed geometries, where the forces on cobalt and all the ammines 
are below the chosen tolerance $0.01 eV/\AA$. 
As for the PESs calculations, we constrain the 
maximum size of a CG step to be $0.001 \AA$ in order to provide a 
large sequence of energies. The resulting total energy can subsequently 
be plotted as a function of the Co-N bond length averaged over the six ligands. 
In the case of JT distortions, the average Co-N bond length in the 
axial direction, $r_{ax}$, and the average Co-N bond length in the 
equatorial plane, $r_{eq}$, are more reasonable choices.
\end{itemize}

\subsection{\label{subsection3.2}Calculation of potential energy surfaces}

\begin{figure}
\includegraphics[height=8.5cm,angle=-90]{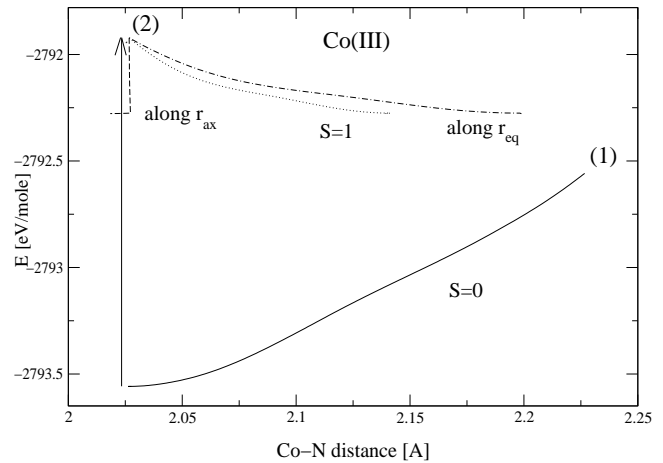}\\
\caption{\label{co3} PESs of Co(III) for spin S=0 (solid line) and S=1 [$\hbar$] (dotted line) 
 as a function of the average Co-N bond length. The numbers (1) and (2)
 indicate the starting points of the CGO. Since Co(III,S=1) becomes JT distorted, 
 we also plot the PES along the axial, $r_{ax}$ (dashed line), and 
 equitorial, $r_{eq}$ (dashed-dotted line), average bond length. 
 The vertical arrow indicates the optical excitation discussed in the text.}
\end{figure}

\begin{figure}
\includegraphics[height=8.5cm,angle=-90]{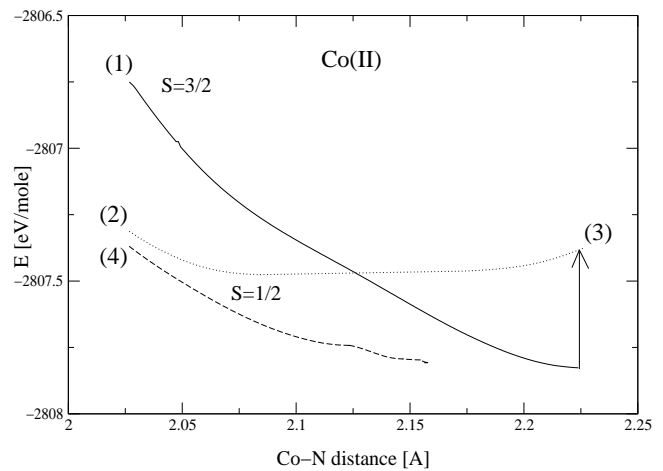}\\
\caption{\label{co2} PESs of Co(II) for spin S=3/2 [$\hbar$] (solid line) and 
 S=1/2 (dotted and dashed lines) as a function of the average Co-N bond length.
 The numbers (1) to (4) indicate the starting points of the CGO. 
 The dotted line shows the non-JT distorded PES, while the dashed line shows a
 JT distorted PES of Co(II,S=1/2). The vertical arrow indicates the optical excitation
 discussed in the text.}
\end{figure}

In the following we describe the procedure of how we obtained the 
PES plots in figures \ref{co3} and \ref{co2}. 

Figure \ref{co3} shows the ground state (S=0) and excited state (S=1) PESs of Co(III). 
For the ground state PES we started a CGO at the Co(II,S=3/2) EQGEO, denoted as starting point (1),
but with the spin fixed to S=0 and charge +3e. For the excited state PES (S=1) we started CGOs at both the 
EQGEOs of Co(III,S=0), starting point (2), and Co(II,S=3/2) (not shown for clarity). 
The vertical arrow indicates the optical excitation
energy from the Co(III,S=0) EQGEO, where it is assumed that the excited state is initially not JT 
distorted, since the excitation is almost instantaneous.
The dotted line in figure \ref{co3} corresponds to total energy versus the 
average Co-N bond length. Since Co(III,S=1) is JT unstable, we also plot the total energy versus the 
average axial Co-N bond length, $r_{ax}$, and the average equatorial Co-N bond length, $r_{eq}$, shown
by dashed lines. One can see that essentially only $r_{eq}$ changes 
when starting a CGO from the Co(III,S=0) EQGEO. 

In order to obtain smooth PESs it is essential to treat 
the ammines as rigid bodies. If one relaxes all the atoms, the resulting PES would be much more 
complicated, i.e. the PES would display a sequence of short parabolic curves stemming 
from periods of contracting or stretching the N-H bond lengths, or from changing the 
H-N-H angles. The resulting PES would not resemble smooth parabola-like curves 
(at least for the non-JT distorted ground states) anymore, since the 
reaction coordinate would not simply be the change of the Co-N bond length. Although 
equilibrium energies of the constraint complexes might be higher than the all-atom-relaxed 
ones, this is not a problem, since we are interested in energy differences and finally 
in the sum of the Co(II) and Co(III) system energies (overall energy shift which does not affect the rate). 

Figure \ref{co2} shows the ground state S=3/2 and excited state 
S=1/2 PESs of Co(II) as a function of the Co-N bond length 
averaged over all six ligands. Using an analogue procedure, 
the Co(II,S=3/2) PES results from a CGO starting at the 
Co(III,S=0) EQGEO, (1), where we have fixed the spin at S=3/2 and the charge at +2e. 
The dotted line shows the Co(II,S=1/2) PES, which was 
obtained from starting at the non-JT distorted Co(III,S=0) and Co(II,S=3/2) EQGEOs, 
(2) and (3) respectively. They stay non-JT distorted. 
The vertical arrow indicates the optical excitation
to the non-JT distorted Co(II,S=1/2) state. Starting geometry (4) is JT distorted and hence lower in
energy than (1). Further relaxation with the CGO method (dashed curve) 
lowers the energy drastically by up to $7.7 kcal/mol$ compared to the 
non-JT relaxed energy (dotted curve). The non-JT distorted PES (dotted curve) is hence only meta stable.
The energy of the Co(II,S=1/2) EQGEO is only 0.3 kcal/mol 
(less than $k_BT$) higher than the Co(II,S=3/2) EQGEO.

Table \ref{table1} contains calculated equilibrium Co-N bond lengths 
of the ground and excited states and their experimental analogous.
For comparison it also shows the ones of the $Co(H_2O)_6^{2+/3+}$ ground states.
As typical for the DFT method they are overestimated, about 3\% for $NH_3$ 
and 8\% for $H_2O$ ligands. 

Table \ref{table2} compares theoretical and experimental optical excitation energies.
The excitation energy ${}^2E_g \leftarrow {}^4T_{1g}$ for $Co(NH_3)_6^{2+}$ is not known from 
experiment but is very important. If this energy separation is low enough Co(II,S=1/2) is 
well-populated at ambient temperatures and the electron transfer reaction is spin-allowed. Although absolute 
excited state energies from DFT cannot be trusted, 
trends established from DFT are often correct. From table \ref{table2} we can obtain the following.
Since DFT underestimates optical gaps such as the $D_q$ ligand-field, DFT favors high-spin compounds.
For instance, the ligand-field splitting of Co(III) is 274 kJ/mol\cite{mcquarrie}
while SIESTA gives 246 kJ/mol.
In other words, excitations from low-spin to high-spin (${}^3T_{1g}\leftarrow{}^1A_g$) are underestimated 
while excitations from high-spin to low-spin (${}^2E_g \leftarrow {}^4T_{1g}$ and 
${}^4T_{1g} \leftarrow {}^6A_{1g}$) are overestimated. Similar results were found for
singlet-triplet gaps of phenylnitrene and other hypovalent systems\cite{smith}. 

\begin{table}
\caption{\label{table1} Equilibrium Co-ligand bond lengths for cobalt hexaammine and hexaaqua
                        complexes in units of $\AA$. Jahn-Teller (JT) distorted values: 
                        $r_{ax}$ denotes the Co-ligand bond length in axial direction, 
                        while $r_{eq}$ in the equatorial plane.}
\begin{ruledtabular}
\begin{tabular}{cc|c|c}
                         &                 &      calcd. [$\AA$]   &   exptl. [$\AA$] \\ \hline
  $Co(NH_3)_6^{3+}$&${}^1A_g$& 2.03&1.96\cite{geselowitz1}\\  
                   &JT distorted ${}^3T_{1g}$& $r_{ax}=2.01$    &\\ 
                   &                          & $r_{eq}=2.20$    &\\ \hline
  $Co(NH_3)_6^{2+}$&${}^4T_{1g}$&2.22&2.16\cite{herlinger}       \\
                   &${}^2E_g$&2.09-2.17&              \\ 
                   &JT distorted ${}^2E_g$& $r_{ax}=2.42$       &              \\ 
                   &                       & $r_{eq}=2.03$       &              \\ \hline\hline
  $Co(H_2O)_6^{3+}$&${}^1A_g$&2.02&1.87\cite{richens}\footnote{p.447}\\  
  $Co(H_2O)_6^{2+}$&${}^4T_{1g}$&2.25&2.08-210\cite{richens}\footnote{p.441}\\
\end{tabular}
\end{ruledtabular}
\end{table}

\begin{table}
\caption{\label{table2} Spin-forbidden transition energies in units of $10^{3}cm^{-1}$.}
\begin{ruledtabular}
\begin{tabular}{c|c|c|c}
  ${}^3T_{1g}\leftarrow {}^1A_g$ & $Co(H_2O)_6^{3+}$ & $Co(NH_3)_6^{3+}$&low-spin\\ \cline{1-3}
               calcd. &$7.15$&$13.23$&to\\  
               exptl. & ($> 1.89$\cite{navon})\footnote{from Cobalt-59 NMR fraction of high-spin $< 10^{-4}$. 
               This is not an order of magnitude estimate, but a true upper bound set by experimental resolution.}
               & $13.70$\cite{wilson}\footnote{corresponds to the maximum of the optical absorption peak at $T = 8 K$ 
which stems from a vertical transition in line with the classical Franck-Condon principle.}&high-spin\\ \hline\hline
  ${}^2E_g \leftarrow {}^4T_{1g}$& $Co(H_2O)_6^{2+}$ & $Co(NH_3)_6^{2+}$&high-spin\\ \cline{1-3}
               calcd. &  $8.46$ &$3.68$&to\\  
               exptl. &  $\approx 6.4$\cite{gailey}\footnote{from a fit of a four parameter octahedral ligand-field theory (Dq, B, C, $\lambda$) including spin-orbit coupling to a circular dichroism spectrum of 
               $[Zn(H_2O)_6]SeO_4]$ doped by $Co(H_2O)_6^{2+}$ at $T = 80 K$. 
               Ligand-field theory is rather good for 
               electronegative ligands such as oxygen.}& $-$ &low-spin\\ \cline{1-3}
  ${}^4T_{1g} \leftarrow {}^6A_{1g}$& $Fe(H_2O)_6^{3+}$ & \\ \cline{1-3}
               calcd. &$19.01$&&\\ 
               exptl. & $14.22$\cite{hammes}\footnote{from ions doped into beryl crystal}&&\\
\end{tabular}
\end{ruledtabular}
\end{table}

\begin{table}
\caption{\label{table3}Energy difference $\Delta E^{3+(2+)}(r)$ in units of $10^3 cm^{-1}$ at the equilibrium
position $r_{3+(2+)}^0$ of Co(III,S=0) (Co(II,S=3/2)) and at the transition state $r^\dagger$ of the conventional
spin-forbidden process. The values of $r^\dagger$ of this work, Larsson's, and Newton's are 2.09, 2.03, and 2.14 $\AA$, respectively.}
\begin{ruledtabular}
\begin{tabular}{c|c|c|c|c|c|c}
&&$r_{2+}^0$&$r^\dagger$&&$r_{3+}^0$&$r^\dagger$\\
\hline 
this work\footnote{DFT-GGA method, the $\Delta E^{2+}$ value is w.r.t. non-JT distorted Co(II, S=1/2).}&$\Delta E^{3+}$&13,23&9.09&$\Delta E^{2+}$&3,68&-1.19\\
Larsson\footnote{INDO-CI method, empirically corrected energies by 
Eqs. (\ref{c1},\ref{c2}) with $c=1,100 cm^{-1}$, energy difference $\Delta E^{3+}(r^\dagger)$ 
is estimated from their Fig. 5}\cite{larsson}
&&13,7&12.0&&3,1-4,1&$\lesssim 0$\\
Newton\footnote{UHF+UMP2 method, empirically corrected energies by 
Eqs. (\ref{c1},\ref{c2}) with $c=-6,000 cm^{-1}$}\cite{newton91}
&&13,7&8,8&&9,1&5,3
\end{tabular}
\end{ruledtabular}
\end{table}

Unfortunately, in the case of low-spin to high-spin excitations there is 
only one reliable value for Co(III) and no
experimental value available for $Co(H_2O)_6^{3+}$ in order to confirm the trend. 
Nevertheless, we can easily see that the
calculated gap for $Co(H_2O)_6^{3+}$ ($7,150 cm^{-1}$) has to be a lower bound analogous to
Co(III). The argument is as follows. From ligand-field theory 
the excitation energy ${}^3T_{1g}\leftarrow{}^1A_g$ is additive in the 
ligand field $D_q$, which is about 25\% larger for $NH_3$ compared to $H_2O$\cite{larsson}. Taking
a value $D_q=274 kJ/mol$ for Co(III) \cite{mcquarrie} and $200  kJ/mol$ for 
$Co(H_2O)_6^{3+}$ \cite{johnson},
we can correct the excitation energy of Co(III), $13,700 cm^{-1}$, by the $D_q$-difference and  
obtain an approximate value for $Co(H_2O)_6^{3+}$, $7,500 cm^{-1}$. This is clearly higher
than the calculated value $7,150 cm^{-1}$.
Having this trend established we can conclude that the excitation ${}^2E_g \leftarrow {}^4T_{1g}$ of
$Co(NH_3)_6^{2+}$, $3,680 cm^{-1}$, is an upper bound on the real value. 

A further confirmation of the quality of our excitation energy of Co(II) 
comes from the fact that our value is similar to Larsson's with INDO-CI ($3,100-4,100 \text{cm}^{-1}$ 
after applying correction Eq. (\ref{c2})\cite{larsson}). 
According to figure \ref{co2}, the JT distorted Co(II,S=1/2) equilibrium energy is 
only slightly higher than the non-JT distorted Co(II,S=3/2) equilibrium energy.
Since the S=1/2 PESs are likely too high by a constant energy shift, 
Co(II,S=1/2) could be close to degenerate with Co(II,S=3/2)
or even be the true ground state. 
Note that these considerations do not include multiplet-spitting due to spin-orbit
coupling, as well as entropy effects on the energy ($\sim -TS=-k_BTln(g)$, where g is the 
degeneracy of the multiplet).  


In table \ref{table3} we compare the important quantity $\Delta E^{3+/2+}(r)$ from Eqs. (\ref{dE3}/\ref{dE2}) 
for various bond lengths r between this, Larson's\cite{larsson} and Newton's\cite{newton91} work. 
$\Delta E^{3+/2+}(r_{3+/2+}^0)$ are the excitation energies out of the equilibrium states 
Co(III,S=0)/Co(II,S=3/2) to Co(III,S=1)/Co(II,S=1/2), 
while $\Delta E^{2+/3+}(r^\dagger)$ is their energy difference at the 
spin-forbidden transition state. The excited state Co(III, S=1) is always much too high
in energy to be relevant and is not important for further consideration. 
However, the value and sign (!) of $\Delta E^{2+}(r^\dagger)$ decides
whether the process will be spin-forbidden (allowed by weak spin-orbit coupling) ($\Delta E^{2+}>0$),
or whether it is spin-allowed ($\Delta E^{2+}<0$). The former process can be described by second
order perturbation theory\cite{sutin1,newton86,newton91} using Co(II,S=1/2) as an virtually excited
state coupled to at the spin-forbidden TS, 
the latter one is through direct coupling and does not involve spin-orbit coupling\cite{larsson}. 
As one can see our value is $\Delta E^{2+}(r^\dagger)<0$ and hence the reaction along the 
lowest energy pathway is spin-allowed.

\begin{figure}[t]
\includegraphics[width=8.0cm,angle=0]{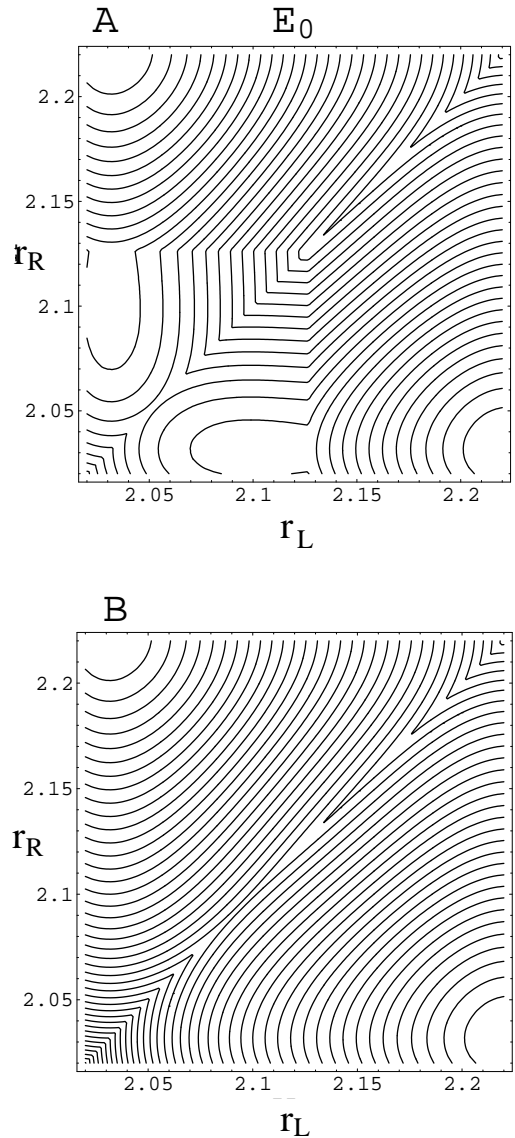}\\
\caption{\label{2PES}Contour plot of the PES of total 2 Co-system with contour level spacing 
0.03 eV = 0.69 kcal/mol. Only results from the non-JT distorted complexes are shown. A)
Example of a spin-allowed reaction using Co(III, S=0) and non-JT distorted Co(II, S=1/2)
for Co(II)-N bond lengths $r_2<2.125 \AA$ and Co(II, S=3/2) for $r_2>2.125 \AA$ (see Fig. \ref{co2}). 
Within the $O_h$ group this is the lowest energy reaction pathway.
B) PES of conventional spin-forbidden reaction originating 
from Co(III, S=0) and Co(II, S=3/2).}
\end{figure}

\begin{figure}[t]
\includegraphics[width=8.5cm,angle=0]{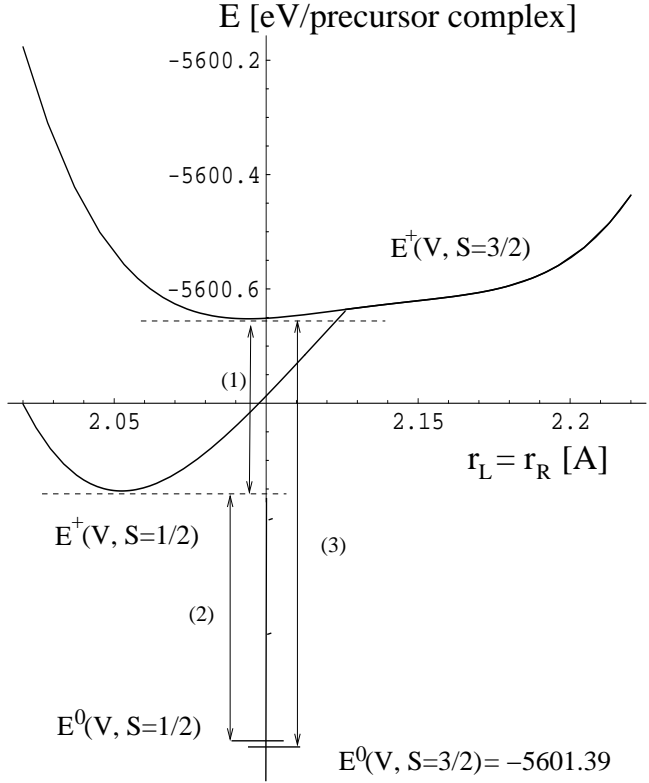}\\
\caption{\label{CoTS}Cross-section of 2Co-PES, written as E(total charge in roman letters, total spin), 
along $r_L=r_R$. The TS is defined as the saddle point.
         Using Co(III,S=0) and non-JT distorded Co(II,S=1/2) lowers the TS by 
         $0.3 eV/\text{precursor complex} = 6.92 kcal/mol$ compared
         to using Co(III) and Co(II,S=3/2) (arrow (1)). Arrows (2) and (3) indicate the two 
activation barriers being the difference between the energy at the transition state (superscript 
$\dagger$) and at the EQGEO (superscript 0).}
\end{figure}

Figure \ref{2PES} shows the PES of the total 2 Co-system, i.e. the sum of the single complex
energies. The minimum in the left top corner at ($r_L=2.03\AA,r_R=2.22\AA$) 
corresponds to equilibrium $Co_L(III)Co_R(II)$, where
the subscripts L and R stand for ``left'' and ``right''
, respectively, and simply distinguish the two complexes. 
The minimum in the right bottom corner at ($r_L=2.22\AA,r_R=2.03\AA$) 
corresponds to equilibrium $Co_L(II)Co_R(III)$. The variables $r_L$ and $r_R$
are the Co-N bond lengths of the left and right complex.
For simplicity we can only show the non-JT distorted complexes, since a single JT distorded
complex depends on two variables, $r_{ax}$ and $r_{eq}$. This would resolve in a total energy
which depends on more than two variables and cannot be plotted.
Part A is plotted using the ground state Co(III,S=0) (solid line in Fig. \ref{co3})
and Co(II,S=1/2) for Co(II)-N bond length $r_2<2.125 \AA$ (dotted line in Fig. \ref{co2}) and
Co(II,S=3/2) for $r_2>2.125\AA$ (solid line in Fig. \ref{co2}). This is an example (by not including 
JT distorted complexes of lower energy) of a spin-allowed reaction. However, if we restricted ourselves
to the $O_h$ group, this would be the lowest energy pathway.
Such a reaction starts out at equilibrium $Co_L(II,S=3/2)Co_R(III,S=0)$, 
then activated by thermal fluctuations 
changes to the $Co_L(II,S=1/2)Co_R(III,S=0)$ PES at ($r_L=2.125 \AA, r_R=2.03 \AA$)
through spin-orbit coupling in only first order (presumably adiabatic). 
Coming back to figure \ref{2PES}, part B shows the PES of Co(III,S=0) and
Co(II,S=3/2) and hence describes the conventional spin-forbidden reaction with an energetically
higher transition state.

In the following we have to obtain the activation barrier of the reaction.
In figure \ref{CoTS} the transition states for spin-allowed (lower graph) and
spin-forbidden (upper graph) are plotted. The graphs are the cross-section of the 2Co-PES 
from figures \ref{2PES} A and B along the diagonal 
$r:=r_L=r_R$, i.e. $E(r_L=r,r_R=r)=E_{Co(III)}(r)+E_{Co(II)}(r)$, where 
the Co-N bond lengths $r_L$ and $r_R$ are the bond lengths of the 
``left'' and ``right'' complex. The transition state is defined as the saddle point.
Using non-JT distorted Co(II,S=1/2) leads to a lowering of the spin-forbidden
transition state by $0.3 eV=6.92 kcal/mol$. The spin-allowed and spin-forbidden TSs
are at $r=2.05$ and $2.1 \AA$, respectively.
In the next section we use the spin-allowed activation barrier 
to estimate the hexaammine self-exchange reaction.

Finally a few things are important to keep in mind. 
Application of the DFT method to transition metal ions is tricky, 
even more when one is interested in excited states. There are several well-known 
deficiencies one has to consider.
First, it is well known that DFT suffers from insufficiently treating certain 
correlation effects\cite{fulde}. This concerns mainly the localized d-orbitals of cobalt, 
where the large $d-d$ Coulomb interaction introduces local correlations that are not captured properly 
by the GGA functional. 
The second deficiency concerns optical gaps which are generally underestimated\cite{fulde,LDA+U,GW}.
In our case the excited states involve the ligand-field spitting. This, however, is turned
to our advantage by using it to deduce a trend. It is utilized in the next section to
obtain an essential lower bound on the rate constant.  
More importantly, since DFT is a ground state theory, we expect that 
excitation energies obtained from total energy differences ($\Delta$SCF method) 
are rather reliable. Very good results have been obtained for 
optical spin (singlet-triplet)\cite{cramer} 
and charge\cite{massobrio} excitations.    
Furthermore, total energies are expected to be better for strongly $\sigma$-donating 
$NH_3$ than for weakly $\pi$-donating $H_2O$ ligands, 
because the $NH_3$ ligands are less electro-negative and bind more covalently with 
cobalt $e_g$ orbitals.
Covalent molecular-type systems are well described by the DFT method.
Besides, covalency screens the on-site repulsion of the localized d-orbitals 
reducing correlation effects.

\section{\label{section4}Results}

In this section we give an estimate for the hexaammine self-exchange rate
using previously estimated electron transfer parameters and results from last section.
For now, we take the energies obtained from the DFT calculations literally. For instance,
we assume that high-spin Co(II,S=3/2) is the groundstate according to our calculation
(see Fig. \ref{co2}). However, since DFT stabilizes high-spin Co(II,S=3/2) over 
low-spin Co(II,S=1/2) according to our trend, this rate estimate is a lower bound. 
This is because we first have to thermally excite to the reaction intermediate, Co(II,S=1/2),
which lowers the rate. With regard to our trend, Co(II,S=1/2) could be the true groundstate.

The initial rate constant to the intermediate state  
Co(II,S=1/2)Co(III,S=0) is denoted by $k_{i}$. 
In thermal equilibrium the return reaction with rate constant $k_{-i}$ back 
to the groundstate Co(II,S=3/2)Co(III,S=0) is equal to $k_i$. From the intermediate state the 
spin-allowed electron transfer can occur with rate constant $k_{et}$ 
\begin{eqnarray}
  Co_L(II,S=3/2)Co_R(III,S=0)&\mathrel{\mathop{\rightleftarrows}\limits^{k_i}_{k_{-i}}}&\nonumber\\
  Co_L(II,S=1/2)Co_R(III,S=0)&\stackrel{k_{et}}{\longrightarrow}&\nonumber\\
  Co_L(III,S=0)Co_R(II,S=1/2)&&.\label{2rates}
\end{eqnarray}
The resulting total rate constant is $k$. Since the two steps from Eq. \ref{2rates} 
are incoherent due to relaxization along the surfaces, the total rate $k$
can equivalently be described as starting out at the thermally populated (with probability P) 
intermediate, $Co_L(II,S=1/2)Co_R(III,S=0)$,
from were the spin-allowed electron transfer can occur to
$Co_L(III,S=0)Co_R(II,S=1/2)$ with rate $k_{et}$. This does not require rate constant $k_i$.
Hence, the total spin-allowed reaction rate constant is  
\begin{equation}
   k\ =\ P\ k_{et}\ =\ \frac{g\ e^{-\beta \Delta G^{*}}}{Z}\ k_{et}\label{k2}
\end{equation}
with $k_{et}$ being the electron transfer rate Eq. (\ref{k1}). 
The prefactor or probability P 
applies in the case that Co(II,S=1/2) is the excited state, which has first to be thermally 
populated. $\Delta G^{*}=\Delta E^{*}-T\Delta S$ is the free energy of excitation and Z is the
partition function. 

There are {\bf two main differences} from previous rate estimates: 
\begin{itemize}
\item Co(II,S=1/2) and Co(II,S=3/2) are {nearly degenerate}, i.e Co(II,S=1/2) is slightly
higher in energy than Co(II,S=3/2) but is probably overestimated. 
In order to get a lower bound of the rate we hypothetically trust the DFT energies and 
calculate $\Delta G^*=0.3kcal/mol+k_BT ln(12/4)=3.1 kcal/mol$ 
from the energy and the entropy differences between Co(II,S=1/2) and Co(II,S=3/2). 
Hence, the prefactor of equation (\ref{k2}) is 0.09
using the degeneracies $g$ from figure \ref{cartoon} and 
neglecting multiplet-splitting.

\item The TS of the Co(II,S=1/2)Co(III,S=0) system is 
{\it lower} than the TS of Co(II,S=3/2)Co(III,S=0). 
The inner-sphere contribution to the activation barrier, 
the energy difference of the 2 Co-system between the TS and equilibrium using Co(II,S=1/2), is only 
$E_{in}^\dagger=0.42 eV=9.69 kcal/mol$. For the equilibrium energy of Co(II,S=1/2), we take the
JT distorted value, for TS we use the non-JT distorted one. This seems reasonable, 
because the acceptor, Co(III,S=0), is non-JT distorted. In order to have sufficient nuclear overlap, 
the donor should be non-JT distorted at TS, too.
\end{itemize}

The other parameters are only slightly modified. This is mainly due to using a different nuclear
frequency for Co(II), i.e. $\nu_2(E_g)$ for Co(II,S=1/2) instead of 
$\nu_2(A_{1g})$ for Co(II,S=3/2):

The {\bf product} ${\mathbf K_{{\mathbf equ}}}\text{\boldmath{$\nu$}}_{{\mathbf eff}}=1.4-3\cdot 10^{11}(Ms)^{-1}$ has been estimated before for
the spin-forbidden process by Sutin \cite{sutin1} and Newton\cite{newton86,newton91}
using a preequilibrium constant $K_{equ}=0.013 M^{-1}$ and an effective nuclear frequency 
$\nu_{eff}=347 cm^{-1}$\cite{newton91}.
Since the case at hand is slightly different (low-spin Co(II,S=1/2)), we redetermine $\nu_{eff}$. 
The average harmonic frequency is given by\cite{newton80,brunschwig82,sutin1}
\begin{equation}
  \nu_{eff}^2=\frac{\nu_{solv}^2E^\dagger_{solv}+\nu_{in}^2E^\dagger_{in}}{E^\dagger_{solv}+E^\dagger_{in}},\label{nu}
\end{equation}
where $\nu_{in}=\sqrt{2}\nu_{2}\nu_{3}/\sqrt{\nu^2_{2}+\nu_{3}^2}$ 
is the reduced inner-sphere nuclear frequency of Co(II) and Co(III). Using
the experimental values $\nu_2(E_g)=255 cm^{-1}$ and $\nu_3(A_{1g})=494 cm^{-1}$\cite{buhks}, 
respectively we obtain $\nu_{in}=320 cm^{-1}=9.6\cdot 10^{12}s^{-1}$. 
For the solvent, we use a typical value $30 cm^{-1}=0.9\cdot 10^{12}s^{-1}$\cite{brunschwig82}. 
We do not attempt to estimate $\nu_{2/3}$ from the PESs due to
complications from JT distortions in the case of Co(II,S=1/2). 
The outer-sphere, $E_{out}^\dagger=28/4kcal/mol=7kcal/mol$, was previously
determined by Buhks and colleges applying the Marcus-Levich continuum model for the 
solvent\cite{buhks,newton86}. We neglect entropy contributions, which are quite small for water near 
room temperature\cite{newton80} and vanish for the inner-sphere system, 
if the (harmonic) vibrations are the same for the activated complex and the reactants\cite{sutin1}.
Using these parameters, one obtains $\nu_{eff}=245 cm^{-1}=7.3\cdot 10^{12}s^{-1}$ and 
$K_{equ}\nu_{eff}=9.5\cdot 10^{10}(Ms)^{-1}$.

The {\bf electronic transmission coefficient} for the 
$e_g$-transfer was determined by Larsson and colleges 
to be weakly adiabatic, $\kappa_{el}\approx 0.5$\cite{larsson}, 
using parameters for the spin-forbidden process and
standard Landau-Zener theory\cite{LandauZener}. 
Extended H\"uckel theory was applied to calculate the 
donor(D)-acceptor(A) electronic coupling, $H_{DA}={}_D\!\!\!<\!\!\!e_g|H|e_g\!\!\!>_A$,
for different precursor complexes (apex-to-apex, apex-to-edge, apex-to-side). A subsequent 
orientational averaging gave $\bar H_{DA}\approx 200 cm^{-1}$. 
Their statistical analyses assumes that each complex can rotate 
independently and that each configuration
covers the same solid angle. The Co-Co separation was chosen $7\AA$ (van der Waals contact between
first solvation shells), the Co-N distances of both complexes were fixed at $2.06\AA$ reasonably close
to our TS value, $2.05\AA$. A similar procedure was applied by Newton\cite{newton91}.
Utilizing $\bar H_{DA}$ from above, our parameters $E_{solv}^\dagger$, $E_{in}^\dagger$, $\nu_{in/eff}$ 
and Eqs. (7-11) from Ref.\cite{newton91,errata}, we redetermine the electronic transmission 
and nuclear tunneling factors and obtain adiabaticity $\kappa_{el}=0.73$ and weak nuclear
tunneling $\Gamma_n=1.9$, respectively. 

Finally, the {\bf activation barrier} has inner-sphere and outer-sphere contributions and is lowered
by the average electronic coupling at TS, i.e.
$G^\dagger\approx E^\dagger=E_{in}^\dagger+E_{solv}^\dagger-\bar H_{DA}$,
and is given by $G^\dagger=0.70 eV=16.0 kcal/mol$. This leads to a Boltzmann factor
$e^{-\beta G^\dagger}=2.1\cdot 10^{-12}$. Including all the calculated charge transfer
parameters in Eq. (\ref{k2}) gives a rate constant of $6\cdot 10^{-3} (Ms)^{-1}$, which is about 
2 orders of magnitude larger than experiment. Possible sources of errors are discussed in section
\ref{section5}. 

In the following, we again want to stress the main differences between our treatment
of a spin-allowed process and the conventional spin-forbidden reaction 
by Buhks {\it et al.} and Newton\cite{newton91}.
The main difference is that in our case the low-spin Co(II,S=1/2) complex has a lower energy
than the high-spin Co(II,S=3/2) near the TS as opposed to Buhks {\it et al.} 
and Newton. In their case, the spin excited state Co(II,S=1/2) is much higher
in energy and is only virtually coupled to by weak spin-orbit coupling. This leads to extremely
small rates of order $10^{-10}(Ms)^{-1}$. 
In table \ref{table4} we show for further illustration the individual parameters used to calculate the 
rate according to Eqs. (\ref{k1}) and (\ref{k2}) and compare to Newton's spin-forbidden and hence diabatic 
groundstate reaction and thermally excited adiabatic pathway. The spin-forbidden reaction is significantly
lowered by the weak spin-orbit coupling ($\kappa_{el}<\!\!<1$), while the excited alternative
is thermally not accessible and leads to even smaller rates. 
The ${}^2E_g \leftarrow {}^4T_{1g}$ excitation energy is as large as 
$9,100 cm^{-1}$, while in our case it is only $3,680 cm^{-1}$.

\begin{table*}
\caption{\label{table4}Rate parameters (see Eqs. (\ref{k1}) and (\ref{k2})) used in this (spin-allowed) and Newton's(spin-forbidden and thermally
excited) work. The rates are compared with experimental estimates. (!) indicates main differences
between Newton's and our work.}
\begin{ruledtabular}
\begin{tabular}{c||c|c|c|c|c|c||c}
&g/Z&$K_{equ}$&$\nu_{eff}$[$cm^{-1}$]&$\kappa_{el}$&$\Gamma_n$&$\Delta G^\dagger$[$kcal/mol$]&rate $k$[$(Ms)^{-1}$]\\
\hline
exp.\cite{geselowitz1,geselowitz2,liquidNH3}&&&&&&&$10^{-7}-10^{-5}$\\
\hline
this work&0.28&0.013&245&0.73&1.9&16.3\footnote{includes $\Delta G^*$ from Eq. (\ref{k2})}&$6\cdot 10^{-3}$\\
\hline
Newton\footnote{individual parameters are presented to the best of our knowledge}\cite{newton91}:&&&&&&&\\
thermally excited&&0.013&245&0.67&2.4&30.7(!)\footnote{due to a large ${}^2E_g \leftarrow {}^4T_{1g}$ excitation energy of $9,100 cm^{-1}$}&$2\cdot 10^{-11}$\\
spin-forbidden&&0.013&347&$10^{-4}$(!)\footnote{due to a spin-orbit reduction factor of $\gamma^2=1.8\cdot 10^{-4}$.}&9&24.4&$4 \cdot 10^{-10}$
\end{tabular}
\end{ruledtabular}
\end{table*}

\section{\label{section5}Discussion}
At this point, there are two main possibilities why there is a 2 order of 
magnitude disagreement between our, $6\cdot 10^{-3} (Ms)^{-1}$, and the experimental rate, 
$5\cdot 10^{-5} (Ms)^{-1}$. Either something is not correct with the 
theoretical estimate, or the experimental rate constant is too small.
In the more likely case that the theory misses some details, the sources of possible errors are 
the ${}^2E_g \leftarrow {}^4T_{1g}$ excitation energy and hence the prefactor of Eq. (\ref{k2}).
Increasing the energy of the Co(II,S=1/2) PESs w.r.t. Co(II,S=3/2) by a 
constant energy shift $2,560 cm^{-1}$ and hence increasing the excitation energy to $6,240 cm^{-1}$,
produces the experimental value. This, however, is in conflict with the trend 
of excitation energies established from the DFT method. 

Another vague possibility of errors in our theory
could originate from a more complicated cross-over from the JT distorted Co(II,S=1/2) equilibrium
complex to thermally excited non-JT distorted Co(II,S=1/2) near TS. The cross-over could be rather
unlikely, since the symmetry changes. 

On the other hand, possible experimental issues are beyond our realm of knowledge. Although
a difficult experiment, it has been thoroughly studied over decades.
Nevertheless, generally forgotten, neglected or underestimated side-reactions 
lead to an underestimation of the rate constant. In particular,
the existence of high-spin Co(II,S=3/2) is challenged by our analysis. This leads to the question, if 
the assumption of high-spin Co(II,S=3/2) was wrongly made when deriving certain rates of side-reactions.
Having said this, an error of about 2 orders of magnitude in the rate estimate 
is not as bad as it may sound. Due to the activated nature of the reaction, the exponential
dependence of the rate on the activation barrier and excitation energy 
makes it very sensitive to small errors.

\section{\label{section6}Conclusion}
Within Marcus theory of charge transfer the rate constant of the hexaamminecobalt-self-exchange 
reaction was redetermined. We utilized the DFT code SIESTA to calculate Born-Oppenheimer
potential energy surfaces and spin-excitation energies and use previously determined parameters.  
The main differences from former work is the near degeneracy of Co(II,S=3/2) 
and Co(II,S=1/2). Furthermore, we observed a drastic lowering of the activation barrier ($\sim 6.9 kcal/mol$) 
for the reaction pathway involving and Co(III,S=0) and Co(II,S=1/2). This led to
a rate constant of order $6\cdot 10^{-3} (Ms)^{-1}$ which is 2 orders of magnitude faster than experiment.
Possible sources of errors are outlined in section \ref{section5} and having most likely
to do with the neglect of the proper cross-over treatment from Jahn-Teller distorted 
equilibrium Co(II,S=1/2)
to non-Jahn-Teller distorted Co(II,S=1/2) near the transition state. 
The good quality of our energetics involved in the rate constant evaluation is indicated, 
first because our calculated excitation energy $13,230 cm^{-1}$ of Co(III), 
${}^3T_{1g}\leftarrow {}^1A_g$, agrees well with the experimental value $13,700 cm^{-1}$.
Second, our excitation energy $3,680 cm^{-1}$ of Co(II), ${}^2E_g \leftarrow {}^4T_{1g}$, is close
to Larsson's value from INDO-CI ($3,100 - 4,100 cm^{-1}$). This energy is in 
particular important for charge transfer. Unfortunately, the corresponding optical absorption peak 
could not be found experimentally. This may simply be explained by the spin-forbiddeness and 
vibrational broadening because of differences in equilibrium Co-N bond distances and
a Jahn-Teller unstable excited state. On the other hand, our analysis questions the existence of the high-spin
ground state Co(II,S=3/2) and hence can provide an alternative explanation for the absence 
the the absorption peak.

{\bf Acknowledgement.} We would like to thank P. Ordej\'on, E. Artacho, D. S\'anchez-Portal and J. M. Soler
for providing us with their \textit{ab initio} code SIESTA. 
This work at Davis was supported by the U.S. Department of Energy,
Office of Basic Energy Sciences, 
Division of Materials Research, and also received support from 
NSF IGERT ``Nanomaterials in the Environment, Agriculture, and Technology''.

\end{document}